\documentclass[a4paper,fleqn,usenatbib]{mnras}

\usepackage{color}

\usepackage{newtxtext,newtxmath}

\usepackage[T1]{fontenc}
\usepackage{ae,aecompl}
\usepackage{algorithm,algpseudocode,enumitem}

\newcommand\Step{{\sf Step }}

\usepackage{graphicx}	
\usepackage{amsmath}	
\usepackage{amssymb}	

\newcommand{\bx}{\boldsymbol{x}}
\newcommand{\wb}{\widehat{\beta}}
\newcommand{\wf}{\widehat{f}}
\newcommand{\wn}{\widetilde{n}}

\title[Photo-z density estimation under selection bias]{
A unified framework for constructing, tuning and assessing photometric redshift density estimates in a selection bias setting
}

\author[P.~E.~Freeman et al.]{
P.~E.~Freeman,$^{1}$\thanks{E-mail: pfreeman@cmu.edu}
R.~Izbicki$^{2}$
and A.~B.~Lee$^{1}$
\\
$^{1}$Department of Statistics, Carnegie Mellon University, 5000 Forbes Avenue, Pittsburgh, PA 15213, USA\\
$^{2}$Department of Statistics, Federal University of S\~ao Carlos, Brazil\\
}

\date{Accepted XXX. Received YYY; in original form ZZZ}

\pubyear{2016}

\begin{document}
\label{firstpage}
\pagerange{\pageref{firstpage}--\pageref{lastpage}}
\maketitle

\begin{abstract}
Photometric redshift estimation is an indispensable tool of precision 
cosmology. One problem that plagues the use of this tool in the era
of large-scale sky surveys is that the bright galaxies that are 
selected for spectroscopic observation do not have properties
that match those of (far more numerous) dimmer galaxies; thus, 
ill-designed empirical methods that produce accurate and precise redshift 
estimates for the former generally will not produce good 
estimates for the latter. 
In this paper, we provide a principled framework for generating 
conditional density estimates (i.e.~photometric redshift PDFs) that takes into 
account selection bias and the covariate shift that this bias induces. We base 
our approach on the assumption that the probability that astronomers 
label a galaxy (i.e.~determine its spectroscopic redshift) 
depends only on its measured (photometric and perhaps other) 
properties $\bx$ and not on its true redshift. With
this assumption, we can explicitly write down risk functions that allow us
to both tune and compare methods for estimating importance weights (i.e.~the
ratio of densities of unlabeled and labeled galaxies for different values of
$\bx$) and conditional densities. We also provide a method for
combining multiple conditional density estimates for the same galaxy into
a single estimate with better properties. We apply our risk functions
to an analysis of $\approx$10$^6$ galaxies, mostly observed by SDSS,
and demonstrate through multiple
diagnostic tests that our method achieves good conditional density
estimates for the unlabeled galaxies.
\end{abstract}




\section{Introduction}

\label{sect:intro}

Photometric redshift (or photo-z) estimation
is an indispensable tool of precision cosmology.
The planners of current and future large-scale photometric
surveys such as the Dark Energy Survey (\citealt{Flaugher05})
and the Large Synoptic Survey Telescope (\citealt{Ivezic08}), which
combined will observe over one billion galaxies, 
require accurate and precise redshift estimates in order to
fully leverage the constraining power of cosmological probes
such as baryon acoustic oscillations and weak gravitational
lensing. Numerous estimators currently exist that
achieve ``good" point estimates of photo-z
redshifts at low redshifts ($z \lesssim 0.5$), where ``good"
means that photo-z and spectroscopic (or spec-z) estimates 
for the {\em same galaxy} largely match, with only a small
percentage of catastrophic outliers. These estimators are
conventionally divided into two classes: template fitters, oft-used 
examples of which include BPZ (\citealt{Benitez00})
and EAZY (\citealt{Brammer08}), and empirical methods such as
ANNz (\citealt{Collister04}).\footnote{See, e.g.~\cite{Hildebrandt10},
\cite{Dahlen13}, and \cite{Sanchez14}, who compare and constrast numerous
estimators from both classes, and references therein.} 
The former utilize
sets of galaxy SED templates that are redshifted until a best match with
a galaxy's observed photometry is found, whereas the latter utilize
spectroscopically observed galaxies to train machine learning methods
to predict the redshifts of those 
galaxies that are only observed photometrically.

Less well established within the field of photo-z estimation,
however, are methods that (1) produce conditional density estimates
(or error estimates) of individual galaxy redshifts {\em and at the
same time}
(2) properly take into account the discrepancy between the populations of
spectroscopically observed galaxies (roughly closer and brighter)
and those observed via photometry only (farther and fainter).

Regarding point (1): the error distributions of photo-z estimates 
are often asymmetric and/or multi-modal, so that single-number summary 
statistics such as the mean or median are insufficient to describe
their shapes. Furthermore, the use of such statistics leads to biased
estimation of parameters in downstream cosmological analyses 
(e.g.~\citealt{Wittman09}); for instance, 
\cite{Mandelbaum08}
demonstrate that the use of the conditional density estimate
$\wf(z \vert \bx)$ (often denoted $p(z)$ in the astronomical
literature and often called the probability density estimate, or PDF) 
reduces systematic error in galaxy-galaxy weak lensing analyses.
(Here, $\bx$ can represent magnitudes and/or colours and/or
other ancillary information measured for a galaxy.)
Several other works have touted the use of $\wf(z \vert \bx)$ as
well, often as a step towards better estimates of ensemble redshift
distributions (usually denoted $N(z)$) in tomographic studies
(e.g.~\citealt{Cunha09}, \citealt{Ball10},
\citealt{Sheldon12}, 
\citealt{CarrascoKind13}, \citealt{CarrascoKind14}, \citealt{Rau15},
\citealt{Bonnett15}, \citealt{DeVicente16}), and standard methods such
as the aforementioned BPZ, EAZY, and ANNz provide
$\wf(z \vert \bx)$ as an available output.

Regarding point (2): it is a well-established truism that in
large-scale surveys there is {\em selection bias}, 
wherein rare and bright galaxies are preferentially selected for
spectroscopic observation. This bias induces a {\em covariate shift}, 
since the properties of these bright galaxies do not
match those of more numerous dimmer galaxies (see e.g.~Figure
\ref{fig:covshift}). This shift
affects the accuracy and precision of empirical photo-z estimates.
One can mitigate covariate shift by estimating
importance weights $\beta(\bx) = f_{U}(\bx)/f_{L}(\bx)$, 
the ratio of the density
of galaxies without redshift labels to those observed spectroscopically.
For instance, \cite{Lima08} attempt to directly estimate $N(z)$ in a covariate
shift setting with a k-nearest-neighbor-based estimator
of the importance weights,
an estimator since utilized by \cite{Cunha09},
\cite{Sheldon12}, and \cite{Sanchez14}. \cite{Rau15}, who propose a weighted
kernel density estimator for $f(z \vert \bx)$, offer two other
methods for computing the weights (quantile regression forest and
ordinal classification PDF). All these weight estimators feature 
parameters that one must tune for proper performance.
One would generally tune estimators by minimizing an estimate of risk
using a validation dataset, but the authors listed
above skirt the issue of tuning by setting the number of nearest
neighbors a priori, or, in the case of \cite{Rau15}, by utilizing a 
plug-in bandwidth estimate via
Scott's rule (see their equation 24).

In this paper, we describe a principled
and unified framework for 
generating conditional density estimates $\wf(z\vert \bx)$
in a selection bias setting:
specifically, we provide a suite of appropriate risk 
estimators, methods for tuning and assessing models, and diagnostic tests 
that allows one to create accurate density estimates 
from raw data $\bx$.\footnote{
One may find
{\tt R} functions implementing our framework at\\
{\tt github.com/pefreeman/CDESB}.}
In {\S}\ref{sect:pstate}, we define both the
problem and our notation. In {\S}\ref{sect:iw}-{\S}\ref{sect:cde} we
show that if we assume that the probability that a
galaxy is labeled depends only on its photometry and not on its true
redshift, which is a valid assumption within the redshift regime probed by 
shallow surveys such as the Sloan Digital Sky Survey (SDSS; \citealt{York00}), 
we can write down risk functions that allow one to properly tune
estimators of both $\beta(\bx)$ and $f(z\vert\bx)$.
These risk functions also allow us to choose from among competing estimators.
In {\S}\ref{sect:combine} we show how one can combine estimators of 
conditional density to improve upon the results achieved by any one estimator
alone. In {\S}\ref{sect:diag} we provide diagnostic tests that
one may use to determine the absolute performance of conditional
density estimators. In {\S}\ref{sect:sdss} we demonstrate our methods 
by applying them to SDSS data. Finally, 
in {\S}\ref{sect:summary} we summarize our results.
In future works, we will provide methods for variable selection (i.e.~the
selection of the most informative colours, etc., to retain from a large
set of possible covariates) and explore methods in which we relax the
galaxy-labeling assumption stated above.

\section{Problem Statement: Selection Bias}

\label{sect:pstate}

The data in a conventional photometric redshift estimation problem 
consist of covariates $\bx \in \mathbb{R}^d$ (photometric colours
and/or magnitudes, etc.) and redshifts $z$. We have access to two data
samples: an independent and identically distributed (i.i.d.) sample 
$\bx^U_1,\ldots,\bx^U_{n_U}$ consisting of
photometric data without associated labels (i.e.~redshifts),
and an i.i.d.~labeled sample
$(\bx^L_1,z^L_1),\ldots,(\bx^L_{n_L},z^L_{n_L})$ 
constructed from follow-up spectroscopic studies.
(For computational efficiency, in our analyses these datasets are
samples taken from larger pools of available labeled and unlabeled data.)
Our goal is to construct a photo-$z$ conditional density 
estimator, $\wf(z|\bx)$, that performs well when applied to
the unlabeled data (where ``well" can be
defined by its performance with respect to a number of metrics; see 
e.g.~{\S}\ref{sect:diag} for two examples).

\begin{figure}
    \includegraphics[scale=0.37,angle=0]{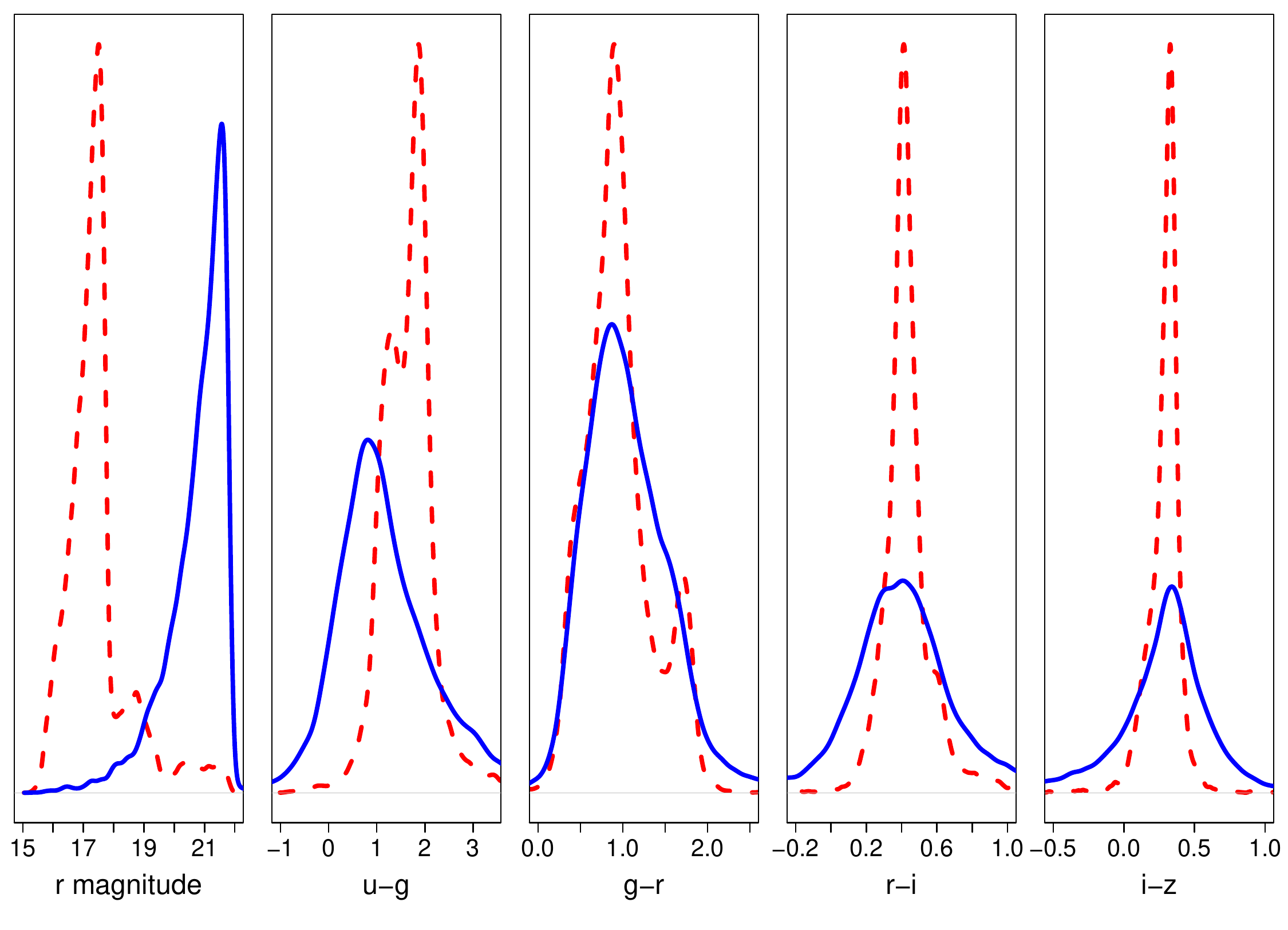}
    \caption{Estimated distributions of $r$-band {\tt model} magnitudes and
    $ugriz$ colours for labeled (i.e.~spectroscopic; red dashed lines) 
    and unlabeled (i.e.~photometric; blue solid lines) galaxies, primarily
    observed by the Sloan Digital Sky Survey (see Section \ref{sect:sdss}). 
    These distributions indicate the bias for selecting brighter galaxies
    for spectroscopic follow-up observations, which induces a covariate
    shift that is here manifested as mismatches in the distributions of
    colours between labeled and unlabeled data.
    }
    \label{fig:covshift}
\end{figure}
 
An issue that arises when constructing $\wf(z|\bx)$
via empirical techniques is that of selection bias.
A standard assumption in statistics and machine learning is that labeled 
and unlabeled data are sampled from similar distributions, which we denote
$\mathbb{P}_L$ and $\mathbb{P}_U$ respectively.
However, as Figure~\ref{fig:covshift} demonstrates, these two distributions 
can differ greatly for sky surveys that mix spectroscopy and photometry; 
brighter galaxies are more likely to be selected for follow-up spectroscopic 
observation.
To model how selection bias affects learning methods,
one needs to invoke additional assumptions about 
the relationship between $\mathbb{P}_L$ and $\mathbb{P}_U$ 
(e.g.~\citealt{Gretton08}, \citealt{MorenoTorres12}). In this work,
we assume that the probability that a galaxy is labeled with a spectroscopic
redshift depends only upon $\bx$ (in accord with
\citealt{Lima08} and \citealt{Sheldon12}); i.e.
\begin{equation}
P(S=1|\bx,z) = P(S=1|\bx) \,
\label{eqn:CSassumption}
\end{equation}
where the random variable $S$ equals $1$ if a datum is labeled and $0$ 
otherwise. This assumption implies covariate shift, defined as
\begin{equation}
f_L(\bx) \neq f_U(\bx), \ \ f_L(z|\bx)=f_U(z|\bx) \,,
\label{eqn:covariate_shift}
\end{equation}
and thus is, as shown below, critical for establishing
the risk function estimators that ultimately allow us to
estimate conditional density estimates $\wf(z \vert \bx)$. 
Following the discussion of Section 2.3 of \cite{Budavari09}, we point out
that assuming $P(S=1|\bx,z) = P(S=1|\bx)$ can be problematic, for instance
when only colours are used in analyses in which the training data are selected
in limited magnitude regimes. In this work we apply our
framework to galaxies at SDSS depth using colours only; for optimal 
performance, one should incorporate those covariates that act in concert
with $z$ to affect selection, e.g., morphology, size, surface brightness,
environment, etc. Nothing in the current framework prevents the 
incorporation of these covariates.

At first glance, it may seem that covariate shift would not pose a problem for 
density estimation; if $f(z|\bx)$ is the same for both labeled and 
unlabeled samples, one might infer that a good estimator of 
$f(z|\bx)$ based on labeled data would also perform well for 
unlabeled data. However, this is generally untrue.
The estimation of $f(z|\bx)$ depends on the marginal 
distribution $f(\bx)$, so an estimator that performs well 
with respect to $f_{L}(\bx)$ may not perform well
with respect to $f_{U}(\bx)$.

We can mitigate selection bias by preprocessing the 
labeled data so as to ensure that sufficient
labeled data lay where the unlabeled data lay.
This allows us
to compute expected values with respect to the distribution $\mathbb{P}_U$
using $\mathbb{P}_L$, similar to the idea of importance sampling in
Monte Carlo methods.
Carrying this mitigation out in practice involves 
two steps: first, we estimate importance weights as a function
of the predictors $\bx$:
\begin{equation}
\beta(\bx) = f_{U}(\bx)/f_{L}(\bx) \,;
\end{equation}
and second, we utilize these weights when estimating
conditional densities $f(z \vert \bx)$.
There are a myriad of estimators both for importance weighting and
for conditional density estimation (see e.g.~\citealt{Izbicki14},
\citealt{Izbicki15});
what we provide here are rigorous procedures 
for tuning their parameters and for choosing among them. 
We note here that our overall procedure can be qualitatively summarized
by the following dictum: one needs good estimates of importance weights at
labeled data points in order to achieve good conditional
density estimates at unlabeled data points. (One can observe how this
dictum plays out mathematically in equation \ref{eqn:cde_risk} below: note
how the importance weight estimates at labeled points enter into it, in
the second term, whereas the importance weight estimates at unlabeled
points do not enter into it at all.)

\section{Importance Weight Estimation}

\label{sect:iw}

A naive method for computing importance weights
$\beta(\bx)$ would involve
estimating $f_U$ and $f_L$ separately and computing the ratio of these
densities, but this approach can enhance errors in individual density 
estimates, particularly in regimes where $f_L \rightarrow 0$ \citep{Sugiyama08}.
Many authors have thus proposed direct estimators of the ratio 
$\beta(\bx)$
(e.g.~\citealt{Sugiyama08}, \citealt{Lima08}, \citealt{Kanamori09},
\citealt{Kanamori12}, \citealt{Loog12}, \citealt{Izbicki14},
\citealt{Kremer15}). As an example, the estimator of \cite{Lima08} and
\cite{Kremer15} is
\begin{equation}
\wb(\bx) = \frac{1}{k} \frac{n_L}{n_U} \sum_{i=1}^{n_U} \mathbb{I}\left( \bx_i^U \in V_k^L \right) \,,
\label{eqn:beta_nn}
\end{equation}
where $V_k^L$ is the region of covariate space containing points that are
closer to $\bx$ then its $k^{\rm th}$ 
nearest labeled neighbor, and
$\mathbb{I}(\cdot)$ is the indicator function.

To choose between importance weight estimators, 
one needs first to optimally tune the parameters of each using the
training and validation data (model selection),
and then assess their performance using the test data (model assessment).
We determine optimal values of tuning parameters by minimizing
a risk function (equation \ref{eqn:beta_risk_int} below).
Generating estimates $\wb(\bx)$ (and by extension
$\wf(z \vert \bx)$, below) implicitly requires smoothing the
observed data, with the smoothing bandwidth set such that estimator
bias and variance are optimally balanced. (See e.g.~\citealt{Wasserman06}.)
For instance,
too much smoothing (e.g.~adopting a value of $k$ that is too large 
in nearest-neighbor-based methods) 
yields estimates with low variance and high bias, i.e.~when applied to
independent datasets sampled from the same distribution, the estimates will
all look similar (i.e.~will have low variance) but will be offset from 
the truth (i.e.~will have high bias).
Too little smoothing (e.g.~$k$ too small) conversely yields high-variance 
and low-bias estimates that overfit the training data. 

\begin{algorithm*}
\caption{Conditional Density Estimation for Photometric Redshifts}
\begin{algorithmic}
\smallskip
\State \textbf{Input:} Pools of labeled (i.e.~spectroscopic) and unlabeled 
(i.e.~photometric-only) data, which we denote $L$ and $U$.\\
\smallskip
\begin{itemize}[leftmargin=1.5cm]
\item [\Step 1.] Randomly sample $n_L$ data from $L$ and $n_U$ data from $U$.
(Here, $n_L = n_U = 15000$.)
\item [\Step 2.] Given number of nearest labeled neighbors $k$ and minimum
number of unlabeled data in neighborhood $u$, repopulate the labeled sample
so as to increase the effective sample size, i.e., to decrease the number of
labeled data with estimated importance weight zero. Such repopulation will
result in photometric covariates distributions that
more closely approximates that of the unlabeled sample.
(See equation \ref{eqn:beta_nn} and Algorithm
\ref{alg:preproc}.)
\item [\Step 3.] Split $L$ and $U$ into training, validation, and test
datasets. (Here, $n_{\rm train} = 7000$, $n_{\rm val} = 3000$, and
$n_{\rm test} = 5000$ for both samples.)
\item [\Step 4.] Apply training and validation data to generate
importance weight esimates $\wb(\bx)$ (via equation \ref{eqn:beta_nn}), 
tuning the estimator so as to
minimize the estimated risk given in 
equation \ref{eqn:beta_risk}.
\item [\Step 5.] Apply importance weights
$\wb(\bx)$ to the estimation of
conditional densities $\wf(z \vert \bx)$ 
(e.g.~equation \ref{eqn:cde_nn}), tuning the
estimator so as to minimize the risk function
given in equation \ref{eqn:cde_risk}.
(Here we apply the conditional density estimators dubbed 
{\tt NN-CS}, {\tt kerNN-CS}, {\tt Series-CS},
and {\tt Combined} by \citealt{Izbicki16}, where {\tt CS} stands
for ``covariate shift" and {\tt Combined} stands for the combination of
the {\tt kerNN-CS} and {\tt Series-CS} estimators. See Section
\ref{sect:combine} for details on combining estimator output.)
\item [\Step 6.] Use the test data to compute the estimated risk 
for each conditional density estimator (equation \ref{eqn:cde_risk}). 
Choose the estimator with the minimum risk.
\end{itemize}
\smallskip
\State \textbf{Output:} Estimated conditional densities
$\{\wf(z\vert\bx_1^L),\ldots,\wf(z\vert\bx_{n_L}^L)\}$ and $\{\wf(z\vert\bx_1^U),\ldots,\wf(z\vert\bx_{n_U}^U)\}$.
\end{algorithmic}
\label{alg:full}
\end{algorithm*}

When estimating importance weights,
we apply the risk function \citep{Izbicki14,Kremer15}
\begin{align}
&R(\wb,\beta) := \int \left(\wb(\bx)-\beta(\bx) \right)^2dP_L(\bx) \nonumber \\
&= \int \wb^2(\bx) dP_L(\bx)-2\int \wb(\bx) \beta(\bx) dP_L(\bx)+\int \beta^2(\bx) dP_L(\bx) \nonumber \\
&= \int \wb^2(\bx) dP_L(\bx)-2\int \wb(\bx) dP_U(\bx)+C(\beta) \,,
\label{eqn:beta_risk_int}
\end{align}
where $dP_U(\bx) = \beta(\bx) dP_L(\bx)$ and
$C(\beta)$ is a term that does not depend on the estimate $\wb$.
(We note that here the calculation of risk is with respect to the labeled
dataset distribution $\mathbb{P}_L$; this is in accord with the dictum stated
above that we need good estimates of importance weights for the labeled 
data in order to achieve good estimates of conditional densities for the
unlabeled data. See equation \ref{eqn:cde_risk} below.)
While we utilize an $L^2$-loss function above
in (\ref{eqn:beta_risk_int}) and below in (\ref{eqn:riskCDE}), one
could in principle substitute other functions based on F-divergences,
log-densities, or notions of $L^1$ loss. However, functions based on
F-divergences and log-densities are overly sensitive to distribution tails
and 
are generally not appropriate for density estimation (see, e.g.~\citealt{Hall87}
and \citealt{Wasserman06}), while estimating a risk based on
$L^1$ loss requires knowledge of the true $f(z \vert \bx)$.
Since in model selection and assessment
we can ignore $C(\beta)$, we rewrite the above as
\begin{align}
J(\wb) &= \int \wb^2(\bx) dP_L(\bx)-2\int \wb(\bx) dP_U(\bx) \,,
\label{eqn:beta_risk_int2}
\end{align}
which we estimate as
\begin{equation}
\widehat{J}(\wb) = \frac{1}{\wn_L}\sum_{k=1}^{\wn_L}\wb^2\left(\boldsymbol{\widetilde{x}}_k^L\right)-\frac{2}{\wn_U}\sum_{k=1}^{\wn_U}\wb\left(\boldsymbol{\widetilde{x}}_k^U\right) \,,
\label{eqn:beta_risk}
\end{equation}
where the tildes indicate that the risk is evaluated using either
validation data (during model selection) or test data (during model assessment).
(Here we use $J$ to represent a risk function in
which the constant term $C(\beta)$ is ignored.)
Among multiple estimators of $\wb(\bx)$, we
choose the one that yields the smallest value of $\widehat{R}$ when applied
to test data.

\section{Conditional Density Estimation}

\label{sect:cde}

Given an estimate $\wb(\bx)$ of the importance weight,
the ratio of densities of the unlabeled and labeled data at the point
$\bx$, our
next step is to compute estimates of the conditional density
$\wf(z \vert \bx)$.
Conditional density estimators include those of \cite{Cunha09} and
\cite{Izbicki15}; see e.g.~\cite{Izbicki16} for more details. To build 
intuition here, we write down the estimator of \cite{Cunha09}, as it is
particularly simple:
\begin{equation}
\wf(z|\bx) \propto \sum_{i \in \mathcal{N}_N(\bx)} \wb\left(\bx^L_i\right) \mathbb{I}\left(z^L_i \in b(z)\right) \,,
\label{eqn:cde_nn}
\end{equation}
where $\mathcal{N}_N(\bx)$ denotes the $N$ neighbors nearest to
$\bx$ among labeled data,
and $b(z)$ denotes the a priori defined bin to which $z$ belongs.
This estimator (up)down-weights labeled data in regions where 
$f_U(\bx)$ is (larger) smaller than $f_L(\bx)$.

In a selection bias setting where $\mathbb{P}_L \neq \mathbb{P}_U$, the 
goal of conditional density estimation is to minimize
\begin{align}
R(\wf,f) := \iint \left(\wf(z|\bx)-f(z|\bx)\right)^2dP_U(\bx)dz \,,
\label{eqn:riskCDE}
\end{align}
i.e.~the risk with respect to the unlabeled data.
Under the covariate shift assumption 
$f_U(z|\bx)=f_L(z|\bx)$, 
one can rewrite the modified risk 
(\ref{eqn:riskCDE}) up to a constant as
\begin{align}
&R(\wf,f) = \iint \wf^2(z|\bx)dP_U(\bx)dz\nonumber \\
&-2 \iint \wf(z|\bx)f(z|\bx)dP_U(\bx)dz + \iint f^2(z|\bx)dP_U(\bx)dz  \nonumber \\
&= \iint \wf^2(z|\bx)dP_U(\bx)dz -2 \iint \wf(z|\bx)\beta(\bx)dP_L(z,\bx) + C(f) \,,
\label{eqn:cde_risk_int}
\end{align}
where the second equality follows from 
$f_U(z|\bx)dP_U(\bx)dz = 
f_L(z|\bx) \beta(\bx) dP_L(\bx) dz=\beta(\bx)dP_L(z,\bx)$. 
Again, this risk depends upon unknown quantities; we ignore $C(f)$ and 
estimate the other terms via the equation
\begin{equation}
\widehat{J}(\wf) = \frac{1}{\wn_U}\sum_{k=1}^{\wn_U}\left[ \int \wf^2\left(z|\widetilde{\bx}_k^U\right)dz \right] -\frac{2}{\wn_L}\sum_{k=1}^{\wn_L}\wf\left(\widetilde{z}^L_k|\widetilde{\bx}^L_k\right)\wb\left(\widetilde{\bx}^L_k\right) \,,
\label{eqn:cde_risk}
\end{equation}
where again the tildes indicate use of validation data in model selection
and test data in model assessment.

\section{Combining Estimators}

\label{sect:combine}

Typical photometric redshift estimation methods utilize one method for
computing $\widehat{z} \vert \bx$ or ${\widehat f}(z \vert \bx)$.
However, one can improve upon the prediction performances of individual
estimators by combining them. Suppose that 
$\wf_1(z|\bx),\ldots, \wf_p(z|\bx)$ 
are $p$ separate estimators of $f(z|\bx)$.  We 
define the weighted average to be
\begin{align}
\wf^{{\boldsymbol \alpha}}(z|\bx)=\sum_{k=1}^p \alpha_k \wf_k(z|\bx),
\label{eqn:comb}
\end{align}
where the weights minimize the empirical risk
$\widehat{R}(\wf^{{\boldsymbol \alpha}})$
under the constraints $\alpha_i \geq 0$ and $\sum_{i=1}^p \alpha_i=1$. 
One can determine 
the solution ${\boldsymbol \alpha}=\left[\alpha_i \right]_{i=1}^p$
by solving a standard quadratic programming problem: 
\begin{align}
\label{eqn:quadratic}
\underset{{\boldsymbol \alpha}:\alpha_i\geq 0,\sum_{i=1}^p \alpha_i=1}{\arg\min} {\boldsymbol \alpha}' \mathbb{B} {\boldsymbol \alpha} -2{\boldsymbol \alpha}'b \,,
\end{align}
where $\mathbb{B}$ is the $p \times p$ matrix 
\begin{align}
\left[\frac{1}{\wn_U} \sum_{k=1}^{\wn_U} \int \wf_i(z|\widetilde{\bx}_k^U)\wf_j(z|\widetilde{\bx}_k^U)dz\right]_{i,j=1}^p
\label{eqn:Bmatrix}
\end{align}
and $b$ is the vector
\begin{align}
\left[ \frac{1}{\wn_L}\sum_{k=1}^{\wn_L}\wf_i(\widetilde{z}^L_k|\widetilde{\bx}^L_k)\wb(\widetilde{\bx}^L_k)\right]_{i=1}^p \,;
\label{eqn:bvector}
\end{align}
the tildes here indicate use of the validation data.

\section{Diagnostic Tests for Estimators}

\label{sect:diag}

Risk estimates, such as those given in equations \ref{eqn:beta_risk} and
\ref{eqn:cde_risk}, allow us to tune estimators and to choose
between estimators, but they do not ultimately convey how well the estimator
performs in an absolute sense. Below we describe
diagnostic tests that one can use to more 
closely assess the quality of different models. Similar tests can be found 
in the time series literature (see, e.g., \citealt{Corradi06}).

\subsection{Assessing uniformity using empirical CDFs}

\label{sect:unif}

Let $\widehat{F}(z|\bx_i)$ denote the estimated conditional 
cumulative distribution function, i.e., let
\begin{equation}
U_i = \widehat{F}(z_i|\bx_i) = \int_0^{z_i} \wf(y \vert \bx_i) dy \,.
\end{equation}
If the chosen estimator performs
well, then the empirical cumulative distribution function (CDF) of the values
$U_1,\ldots,U_n$ will be consistent with the CDF of the
uniform distribution. We can test this hypothesis via, e.g., the
Cram\'er-von Mises, Anderson-Darling, and Kolmogorov-Smirnoff tests.
If the $p$-value output by the test is $> 0.05$,
then we fail to reject the null hypothesis that the data 
$U_1,\ldots,U_{n_{\wb_o}}$ are sampled from a uniform distribution.

\subsection{Assessing uniformity using quantiles}

\label{sect:qq}

We can use the values $U_1,\ldots,U_n$ defined above
to build a quantile-quantile (or QQ) plot, 
by determining the number of data in bins of width $\Delta U$.
Let $J$ be the number of bins, each of which has midpoint
$c_j$ and the fraction of data $\widehat{c}_j$.
The QQ plot is that of the values $\widehat{c}_j$ against $c_j$; if 
the chosen estimators perform well, the points in this plot will approximately 
lie on the line $\widehat{c} = c$.
We assess consistency with uniformity via
the chi-square goodness of fit (GoF) test, which
utilizes the chi-square statistic
\begin{equation}
\chi_{\rm obs}^2 = \sum_{j=1}^{J} \frac{(n\widehat{c}_j - n/J)^2}{n/J} \,.
\end{equation}
We conclude that the difference data are consistent with constancy if the $p$ 
value, the fraction of the time that a value of $\chi^2$ would be observed that
is greater than $\chi_{\rm obs}^2$ if the null hypothesis of constancy is
true, is $> 0.05$. Note that the off-the-shelf GoF test, which allows one to
compute the $p$ value by taking the tail integral of an appropriate 
chi-square distribution, requires
that the number of expected counts in each bin be $\gtrsim$ 5. When that
condition is violated, one should use simulations to estimate $p$ values.

\subsection{Assessing uniformity in interval coverage}

\label{sect:cov}

For every $\alpha_j$ in a grid of values on $[0,1]$ of length $J$, and 
for every observation $i$ in the labeled test sample, 
we determine the smallest interval 
$A_{ij} = [z_i^{\rm lo},z_i^{\rm hi}]$ such that
\begin{equation}
\int_{A_{ij}}\wf(z|\bx_i)dz = \alpha_j \,.
\end{equation}
Then, for every $\alpha_j$, we determine the proportion of
redshifts lie within $A_{ij}$, i.e.~we compute
\begin{equation}
\widehat{\alpha}_j = \sum_{i \in S} \mathbb{I}(z_i^L \in A_{ij}) \,.
\label{eqn:cov}
\end{equation}
If the chosen estimators perform well, then $\widehat{\alpha} \approx \alpha$.
We plot the values of $\widehat{\alpha}$ against corresponding values
of $\alpha$ and assess how close the plotted points are to the line 
$\widehat{\alpha} = \alpha$; we can test for consistency with that line 
using the chi-square GoF test, as described above.

Note that the construction of coverage plots is
also proposed by \cite{Wittman16}, who conclude that the
$\wf(z \vert \bx)$ produced by the template-based
BPZ and EAZY codes are consistently too narrow and approximately correct,
respectively.

\section{Application to SDSS Data}

\label{sect:sdss}

\subsection{Data}

To demonstrate the efficacy of our conditional density estimation method,
we apply it to $\approx$10$^6$ galaxies that are mostly from Data Release 8
of the Sloan Digital Sky Survey \citep{Aihara11}.

To build our unlabeled (i.e.~photometric) dataset, we initially
extract {\tt model} magnitudes for 540{,}235 objects in the sky patch
RA $\in$ [168$^{\circ}$,192$^{\circ}$] and $\delta \in 
[-1.5^{\circ},1.5^{\circ}]$. After filtering these data in the manner
of \cite{Sheldon12}, namely limiting our selection to those data whose
$ugriz$ magnitudes were {\em all} between 15 and 29, and then further limiting
ourselves to data for which
\begin{equation*}
$u < 21$~~{\rm or}~~$g < 22$~~{\rm or}~~$r < 22$~~{\rm or}~~$i < 20.5$~~{\rm or}~~$z < 20.1$ \,,
\end{equation*}
we obtain a sample of 538{,}974 objects.

We use the labeled (i.e.~spectroscopic) dataset of \cite{Sheldon12}
(E.~Sheldon, private communication). This dataset includes 435{,}875 objects
from SDSS DR8 and 31{,}835 objects from eight other sources, or
467{,}710 objects in all. As noted by \citeauthor{Sheldon12}, this
dataset contains a small number of stars. We remove these by excluding
all data with spectroscopic redshift $z_{s} = 0$; after this, we are left 
with 465{,}790 objects.

The steps of our analysis are given in Algorithms \ref{alg:full} and 
\ref{alg:preproc}.
As noted in Section \ref{sect:pstate}, the labeled and unlabeled data that
we analyse are samples from the larger pools of available data described
immediately above. (In this work we set $n_L = n_U = 15{,}000$.) 
This is for computational efficiency,
both from a standpoint of time and memory; 
for instance, if we utilize matrices for storing distances between
data points, we are currently limited to samples of size $\sim$ 10$^4$
when utilizing typical desktop computers.

\begin{algorithm*}[ht]
\caption{Preprocessing Labeled (i.e.~Spectroscopic) Data}
\begin{algorithmic}
\State \textbf{Input:} Number of labeled and unlabeled data, 
$n_L$ and $n_U$.\\
\hspace{0.325in} Labeled dataset $L = \{\bx_1^L,\ldots,\bx_{n_L}^L\}$ and 
unlabeled dataset $U = \{\bx_1^U,\ldots,\bx_{n_U}^U\}$.\\
\hspace{0.325in} Number of nearest labeled neighbors, $k$.\\
\hspace{0.325in} Minimum number of unlabeled data required for selection
of new labeled datum, $u$.
\begin{itemize}[leftmargin=1.5cm]
\item[\Step 1.] Randomly permute all labeled data indices (i.e.~those from the larger pool of all available labeled data).
\item[\Step 2.] Loop over the permuted set of indices $i_p$:\\
estimate $\wb(\bx_{i_p})$ within the volume containing
the $k$ nearest neighbors among $L$ (via e.g.~equation \ref{eqn:beta_nn})\\
if $\wb(\bx_{i_p}) \geq u/k$, accept $\bx_{i_p}$ as a new labeled datum\\
if the number of new labeled data equals $n_L$, terminate loop.
\end{itemize}
\smallskip
\State \textbf{Output:} New labeled dataset $\{\bx_1^L,\ldots,\bx_{n_L}^L\}$.\\
\hspace{0.45in}The importance weights for these data are re-estimated and are applied to conditional density estimation
(e.g.~equation \ref{eqn:cde_nn}, and equation \ref{eqn:cde_risk}).
\end{algorithmic}
\label{alg:preproc}
\end{algorithm*}

\subsection{Data preprocessing: construction of the labeled sample}

\label{sect:preproc}

As shown in equation \ref{eqn:cde_nn}, the estimation of conditional densities
is partially a function of $\wb(\bx^L)$, so there is a distinction to be
drawn between the stated labeled sample size (e.g.~$n_L$ = 15000, drawn from
a pool of size 465{,}790) and the 
effective size (the number of data that contribute to estimation, i.e.~the
number for which $\wb(\bx^L) > 0$). Thus one important step of our method
involves preprocessing the labeled data to increase their effective size.

\begin{figure}
    \includegraphics[scale=0.37,angle=0]{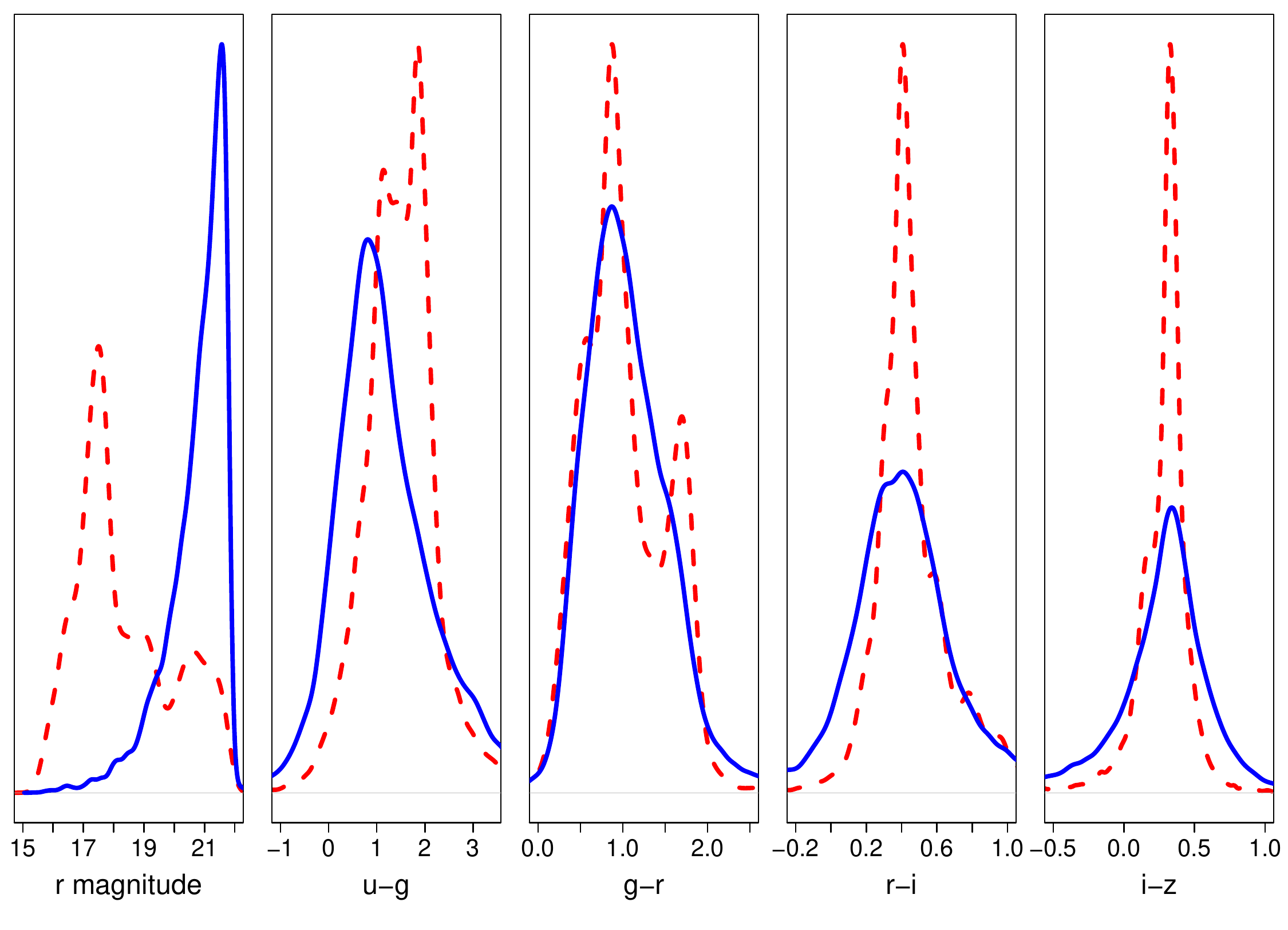}
    \includegraphics[scale=0.37,angle=0]{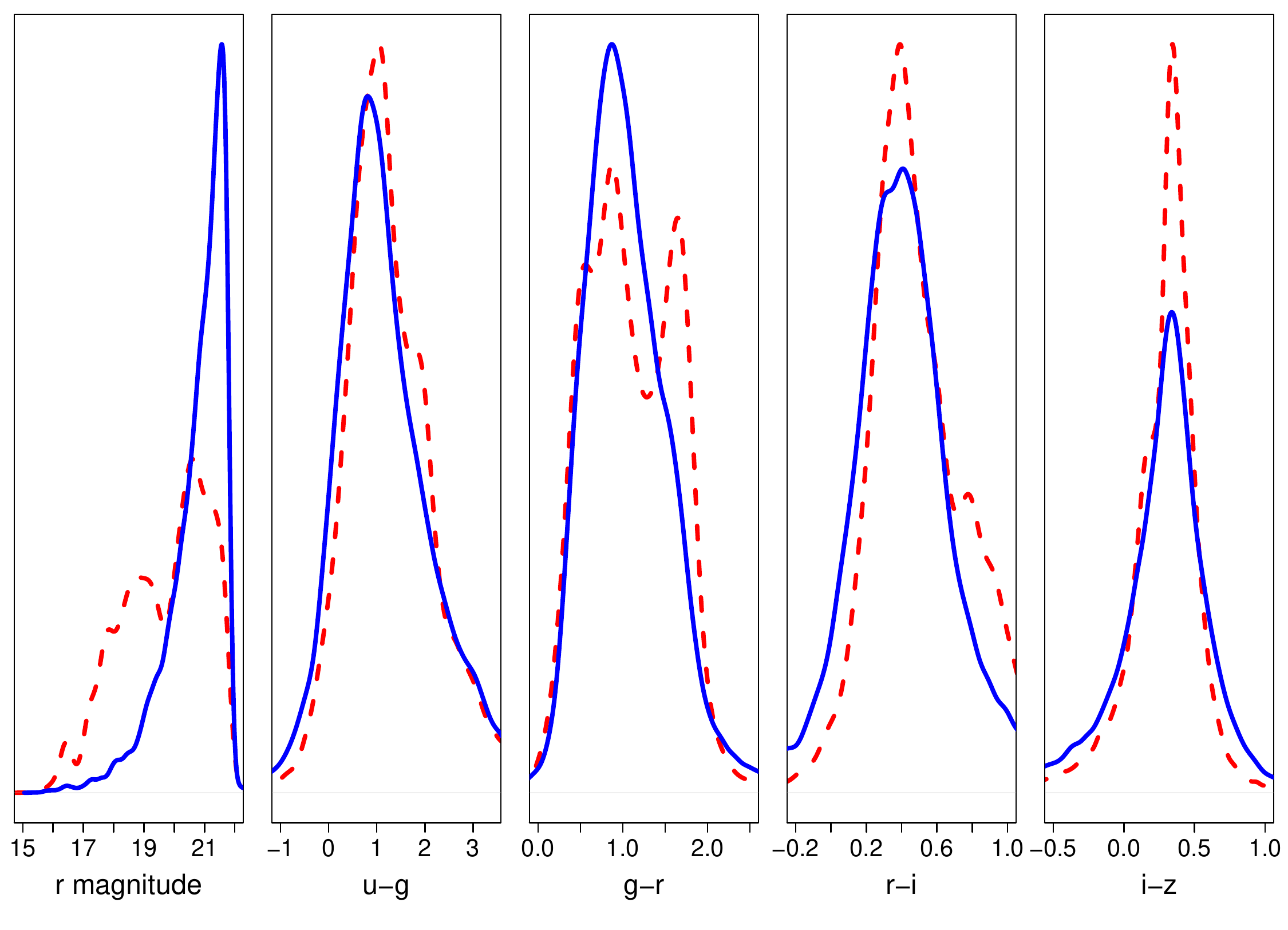}
    \includegraphics[scale=0.37,angle=0]{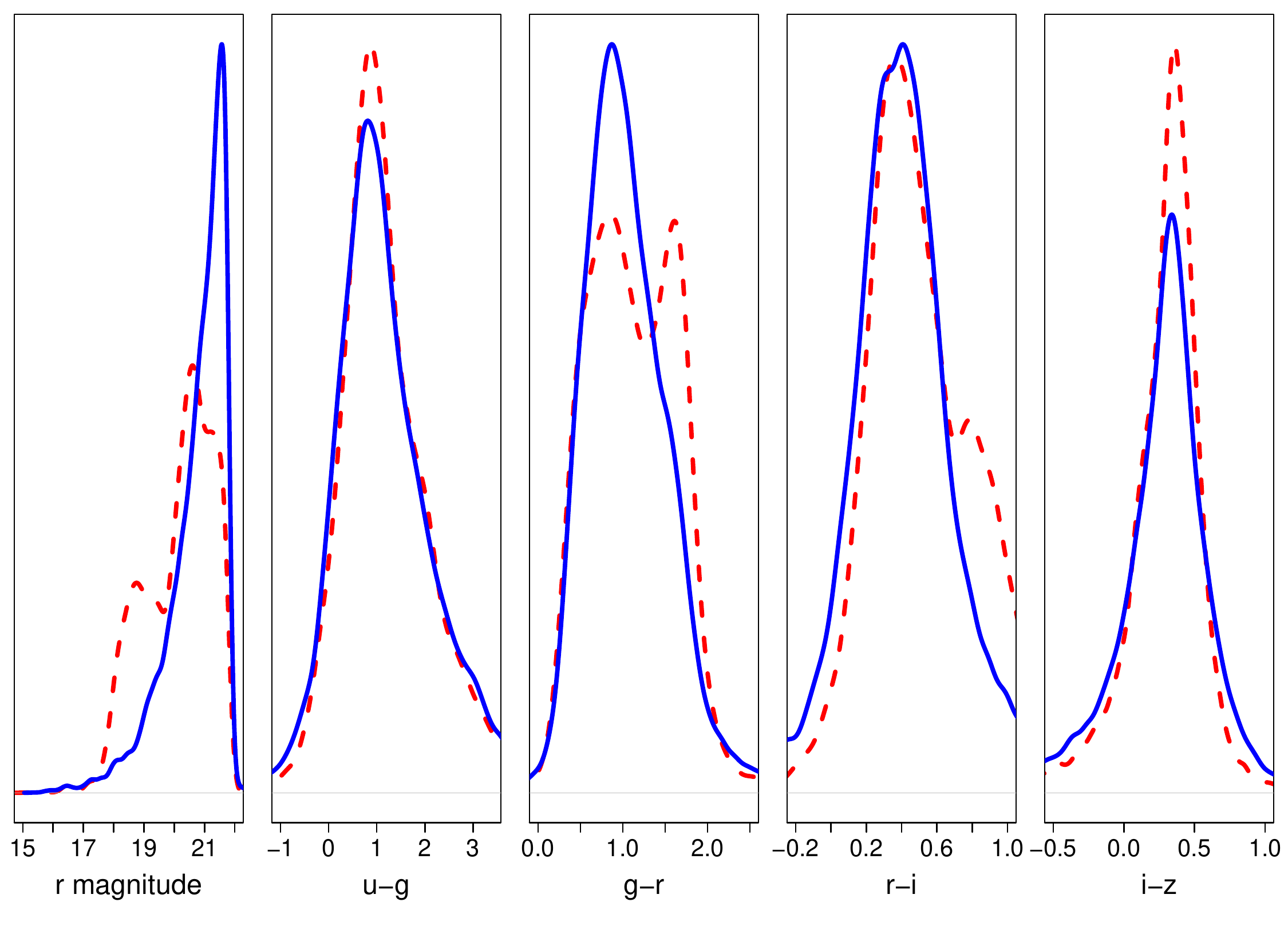}
    \caption{Same as Figure \ref{fig:covshift}, except that here we construct
    the labeled data with the preprocessing scheme outlined in Algorithm
    \ref{alg:preproc} so as to increase the effective sample size (i.e.~the
    number of labeled data with non-zero importance weights). The panels 
    exhibit preprocessing with 
    $\wb_{\rm thr} = 0.1, 0.2$ and $0.3$, respectively from top to bottom
    (see equation \ref{eqn:preproc}). 
    (For larger values of $\wb_{\rm thr}$, we cannot fully populate a
    labeled subsample of size $n_L = 15000$ from the pool of labeled data 
    at our disposal.) In all three cases, the
    distributions for the labeled data (red dashed lines) more closely resembles
    those for the unlabeled data (blue solid lines), relative to the 
    labeled distributions exhibited in Figure \ref{fig:covshift}.
    Subsequent to preprocessing we re-estimate the
    importance weights $\wb(\bx^L)$ and apply these values to the 
    estimation of conditional densities (e.g.~equation \ref{eqn:cde_nn}, and
    equation \ref{eqn:cde_risk}).
    }
    \label{fig:preproc}
\end{figure}

The preprocessing of the labeled data requires the specification of a 
threshold importance weight $\wb_{\rm thr}$. Given $n_L$ and $n_U$
labeled and unlabeled data, respectively, and having specified 
a minimum number of unlabeled data $u$ that would have to lie closer 
to a random labeled point $\bx^L$ (drawn from the larger pool)
than the $k^{\rm th}$ nearest labeled neighbor to that point, we keep
$\bx^L$ as part of our {\em new} labeled dataset if
\begin{equation}
\wb(\bx^L) \geq \wb_{\rm thr} = \frac{u}{k} \,.
\label{eqn:preproc}
\end{equation}
The value $\wb_{\rm thr}$ is not tunable, per se, as different thresholds
yield different labeled datasets, leading to estimated risks that are
not directly comparable. One might conjecture that larger values of
$\wb_{\rm thr}$ are better, in that the distribution of the labeled data
will more closely resemble that of the unlabeled data (see
Figures \ref{fig:covshift} and \ref{fig:preproc}). However, as we
demonstrate below, our results are not
highly sensitive to the choice of $\wb_{\rm thr}$, so long as
$\wb_{\rm thr} > 0$. 

Once preprocessing is complete, we repeat the estimation of the
importance weights $\wb(\bx^L)$ and apply these values when estimating 
conditional densities (e.g.~equation \ref{eqn:cde_nn}, and 
equation \ref{eqn:cde_risk}).

\subsection{Results}

\begin{figure}
    \includegraphics[scale=0.525,angle=0]{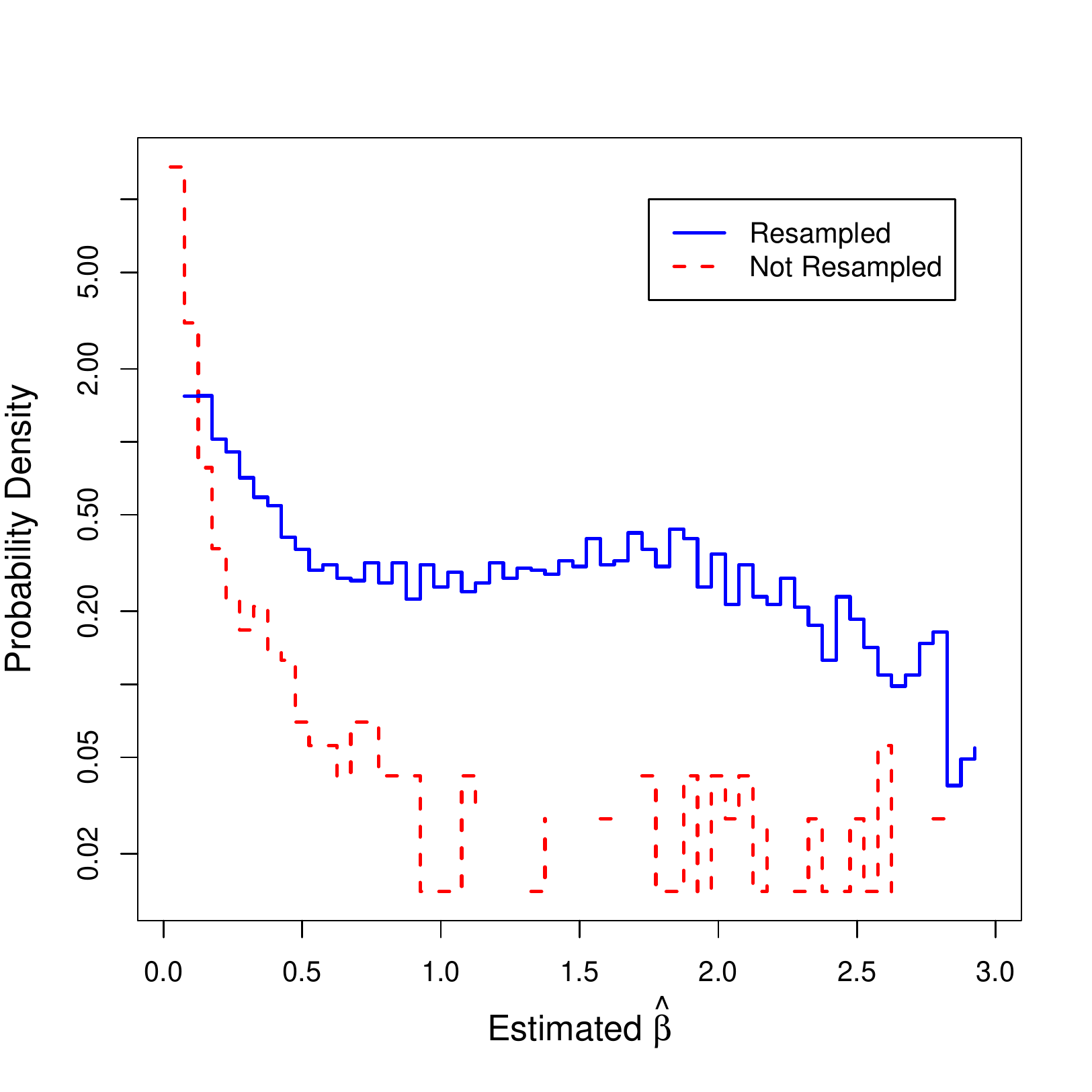}
    \caption{
    Estimated probability density functions for $\wb(\bx^L)$ before
    (red dashed line) and after (blue solid line)
    application of the labeled data preprocessing scheme
    described in Section \ref{sect:preproc} and Algorithm \ref{alg:preproc}.
    Without preprocessing, there are effectively no data with which to
    learn a statistical model linking the covariates and redshift in
    the regime $\wb(\bx^L) \gtrsim 0.5$, a regime that contains 
    $>$ 90\% of the unlabeled data.
    }
    \label{fig:preproc2}
\end{figure}

\begin{figure}
    \includegraphics[scale=0.475,angle=0]{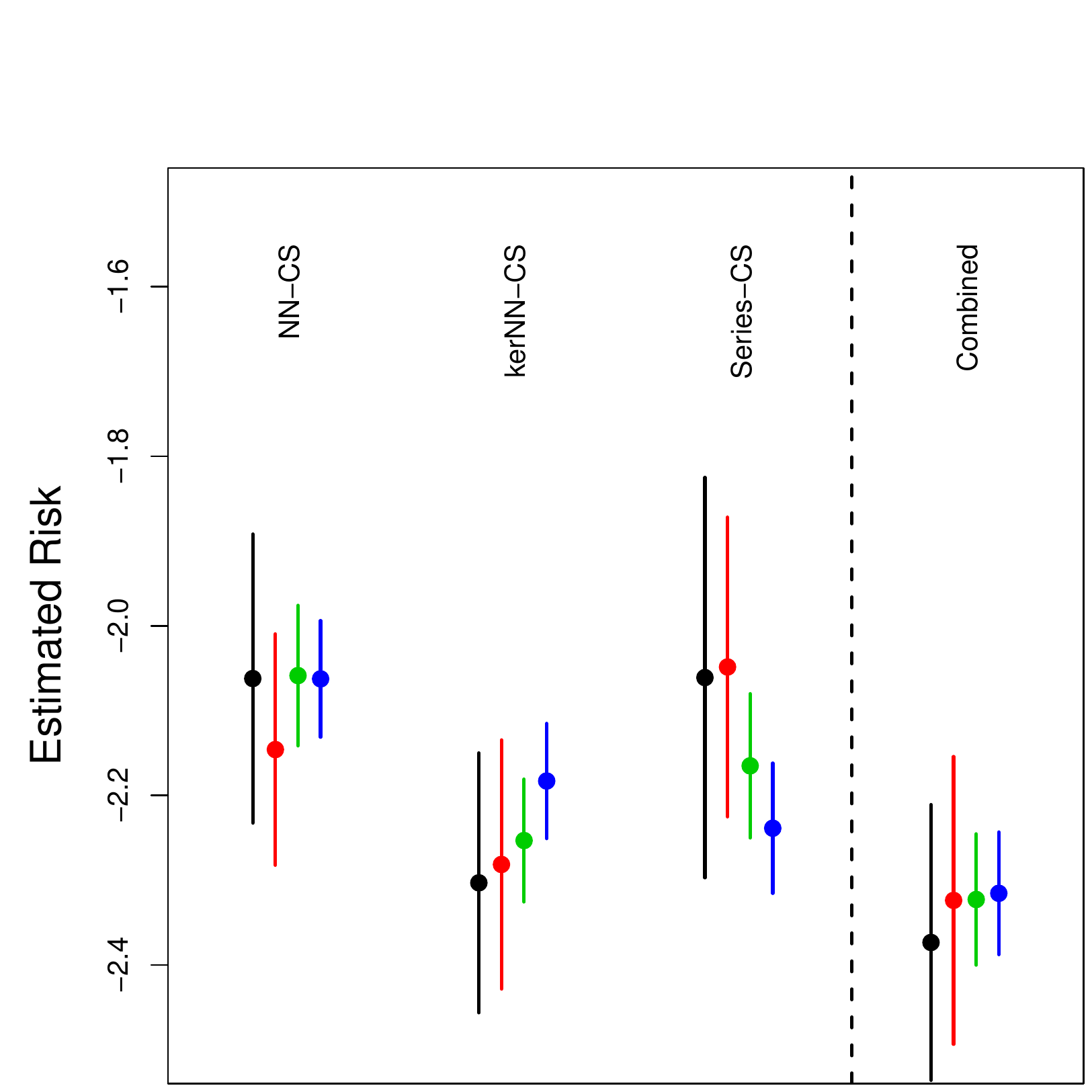}
    \caption{Estimated risk (and standard errors) resulting from the 
    application of the {\tt NN-CS},
    {\tt kerNN-CS} and {\tt Series-CS} conditional density estimators,
    and a combination of the latter two (see Section \ref{sect:combine}).
    The four points in each group represent different values for
    the importance weight threshold $\wb_{\rm thr}$ in our scheme for
    preprocessing labeled data; from left to right, $\wb_{\rm thr} = 0.0,
    0.1, 0.2,$ and $0.3$ (black, red, green, and blue respectively).
    The estimated risks for each $\wb_{\rm thr}$ were adjusted to have 
    mean $-$2.2.
    In all four cases, we observe that combining the {\tt kerNN-CS} and 
    {\tt Series-CS} estimators leads to better conditional density estimates
    than either yield themselves.
    }
    \label{fig:risk}
\end{figure}

Once we generate our new labeled dataset, we split the labeled and 
unlabeled data into training ($n_{\rm train} = 7000$), validation
($n_{\rm valid} = 3000$), and test ($n_{\rm test} = 5000$) sets.
We forego model assessment for importance weight estimators in this work;
\cite{Izbicki16} demonstrate that the estimator given in equation
\ref{eqn:beta_nn} consistently performs better than five other competing
methods over a number of different levels of covariate shift. We apply
equations \ref{eqn:beta_nn} and \ref{eqn:beta_risk} to the training and
validation data to determine the optimal number of nearest neighbors $k$
and importance weights $\wb(\bx^L)$ given $k$.
We then apply these importance weights and the estimated risk in
equation \ref{eqn:cde_risk} to the training and validation data within the
context of three CDE estimators detailed in 
\cite{Izbicki16}: {\tt NN-CS}, the estimator of \cite{Cunha09};
{\tt kerNN-CS}, a kernelized variation of {\tt NN-CS};
and {\tt Series-CS}, the spectral series estimator of
\cite{Izbicki15}.\footnote{
To expand upon the description of the
bias-variance tradeoff in Section \ref{sect:iw}:
We note that any estimates $\wb(\bx)$ and $\wf(z \vert \bx)$
made near (potentially sharp) parameter-space boundaries
will suffer from some amount of additional ``boundary bias." 
Mitigating boundary biases in photo-z estimation is an important
topic that we will pursue in a future work.}
In Figure \ref{fig:preproc2} we demonstrate the importance of preprocessing
to generate the labeled dataset: without it, the construction of 
conditional density estimates for the unlabeled data would be effectively
limited to the regime $\wb(\bx^U) \lesssim 0.5$. The
fraction of unlabeled test set data with $\wb(\bx^U) \leq 0.5$ is
approximately 7.5\%; in contrast, the fraction for $\wb(\bx^U) \leq 3$
is 88.5\%.
In Figure \ref{fig:risk} we demonstrate that for our
particular SDSS data, the {\tt kerNN-CS} estimator on average outperforms the
other two. (Note that the risk can be negative 
because we ignore the positive additive constant $C(f)$; see 
equation \ref{eqn:cde_risk_int}.) We use the method outlined in Section
\ref{sect:combine} to optimally combine the conditional density estimates
from {\tt kerNN-CS} and {\tt Series-CS} and we determine that combining
these estimators, on average, indeed yields better estimates of the
conditional densities $\wf(z\vert\bx)$.

\begin{figure}
    \includegraphics[scale=0.55,angle=0]{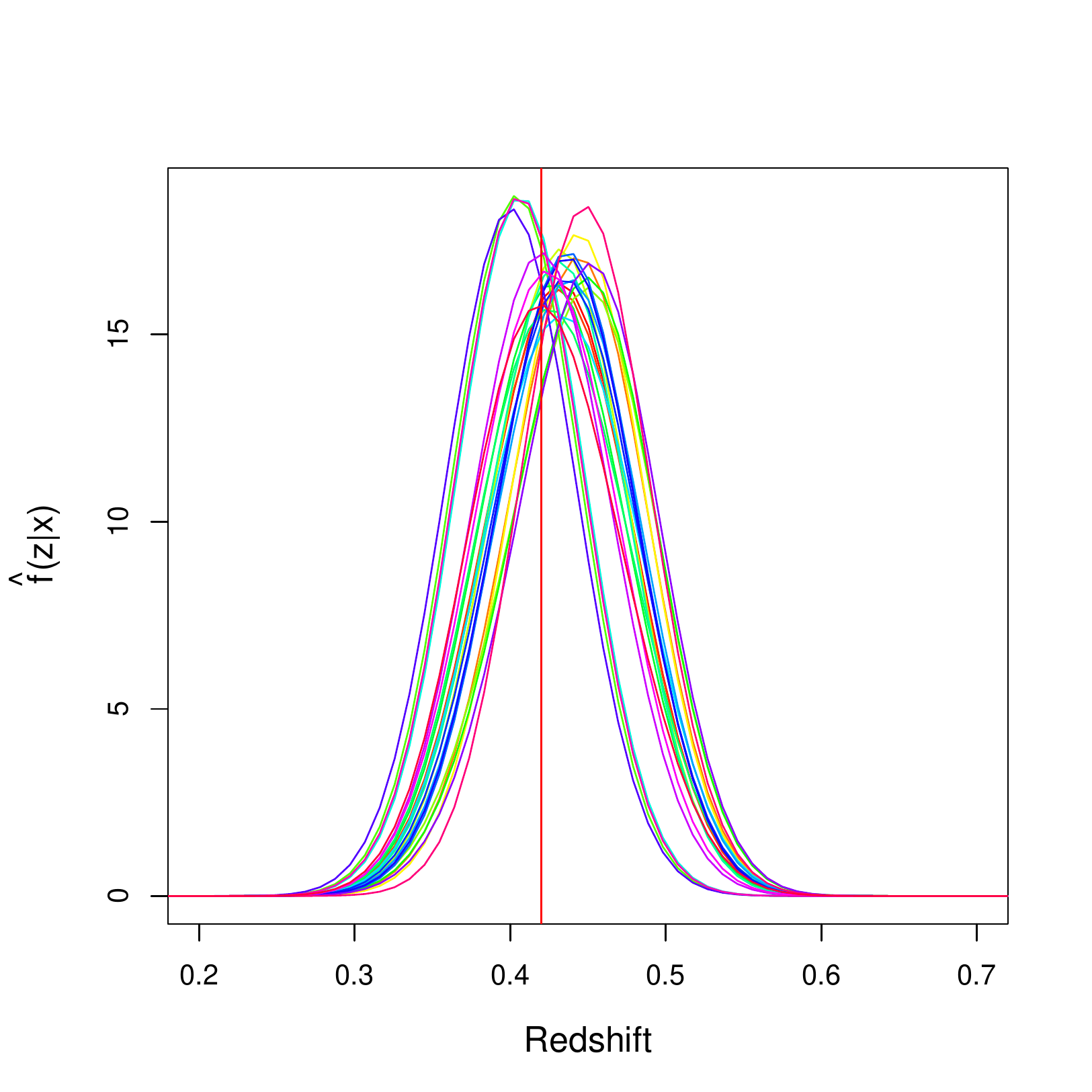}
    \caption{
    Twenty-five estimates of $\wf(z \vert \bx)$ made for a
    single labeled test-set galaxy via bootstrapping of the labeled
    training dataset. The solid red vertical line indicates the true
    redshift. As described in the text, we make a preliminary assessment
    of the noise properties of $\wf(z \vert \bx)$ by computing the
    true redshift quantile for each curve (i.e.~the integral of the curve
    between zero and the true redshift), computing the sample standard
    deviation of these quantiles, and determining the overall mean standard
    deviation. Here, the standard deviation of the quantiles is 
    $\approx$ 0.13; on average the standard deviation is 
    $\approx$ 0.06.
    }
    \label{fig:bootstrap}
\end{figure}

We make a preliminary assessment 
of the noise properties of the estimates $\wf(z\vert\bx)$ as follows. 
We create $n$ bootstrap samples of the 
labeled training data and use them to generate $n$ 
$\wf(z\vert\bx)$ curves for each labeled test datum. 
(See Figure \ref{fig:bootstrap}.) Then we compute
the quantile of the true redshift given each curve, so that for each labeled
test datum we have $n$ quantile values. We use the mean of the standard
deviations of each set of quantile values as a metric of uncertainty.
For our particular SDSS data, we determine this mean uncertainty to be
$\approx$ 0.065. 
We will examine the noise properties of the estimates $\wf(z\vert\bx)$
at greater depth in a future work.

\begin{figure}
    \includegraphics[scale=0.525,angle=0]{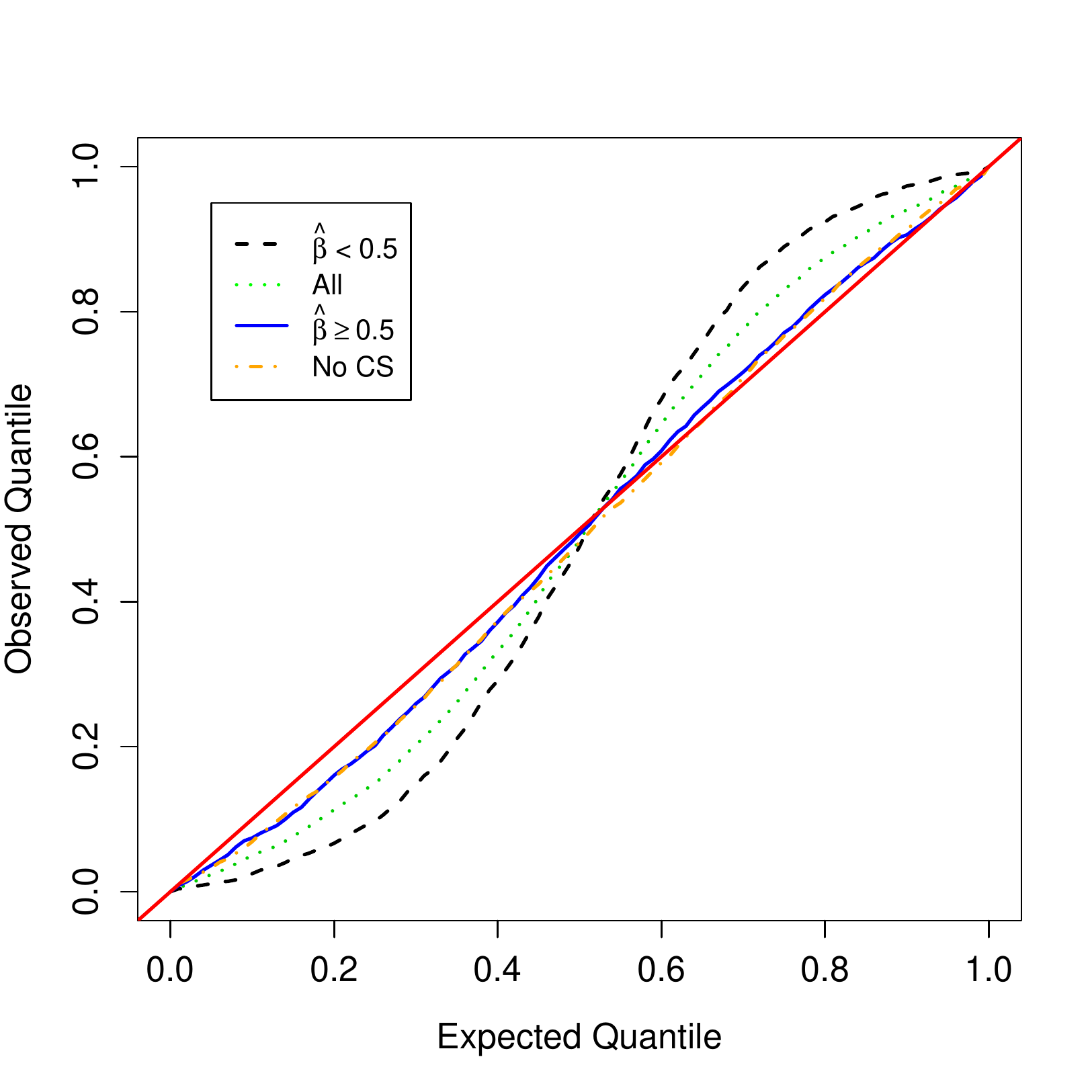}
    \caption{Quantile-quantile results for all labeled test
    data (green dotted line), labeled test data for which 
    $\wb(\bx^L) < 0.5$ and $\geq 0.5$ (black dashed and blue solid lines
    respectively), and labeled test data for the case where we do not mitigate
    covariate shift by incorporating importance weights into estimation
    (orange dashed-dotted line, largely following
    the blue solid line), which effectively
    limits modeling to the regime $\wb(\bx^L) \lesssim 0.5$ (see Figure
    \ref{fig:preproc2}). This plot demonstrates that mitigating covariate
    shift leads to worse conditional density 
    estimates in the portion of covariate space 
    where $\wb(\bx^L) \lesssim 0.5$ and where $\lesssim$ 8\% of the unlabeled 
    data are located (orange line vs. black line), but yields good estimates
    throughout the remaining covariate space (blue line vs. orange line).
    }
    \label{fig:qqall}
\end{figure}

\begin{figure}
    \includegraphics[scale=0.475,angle=0]{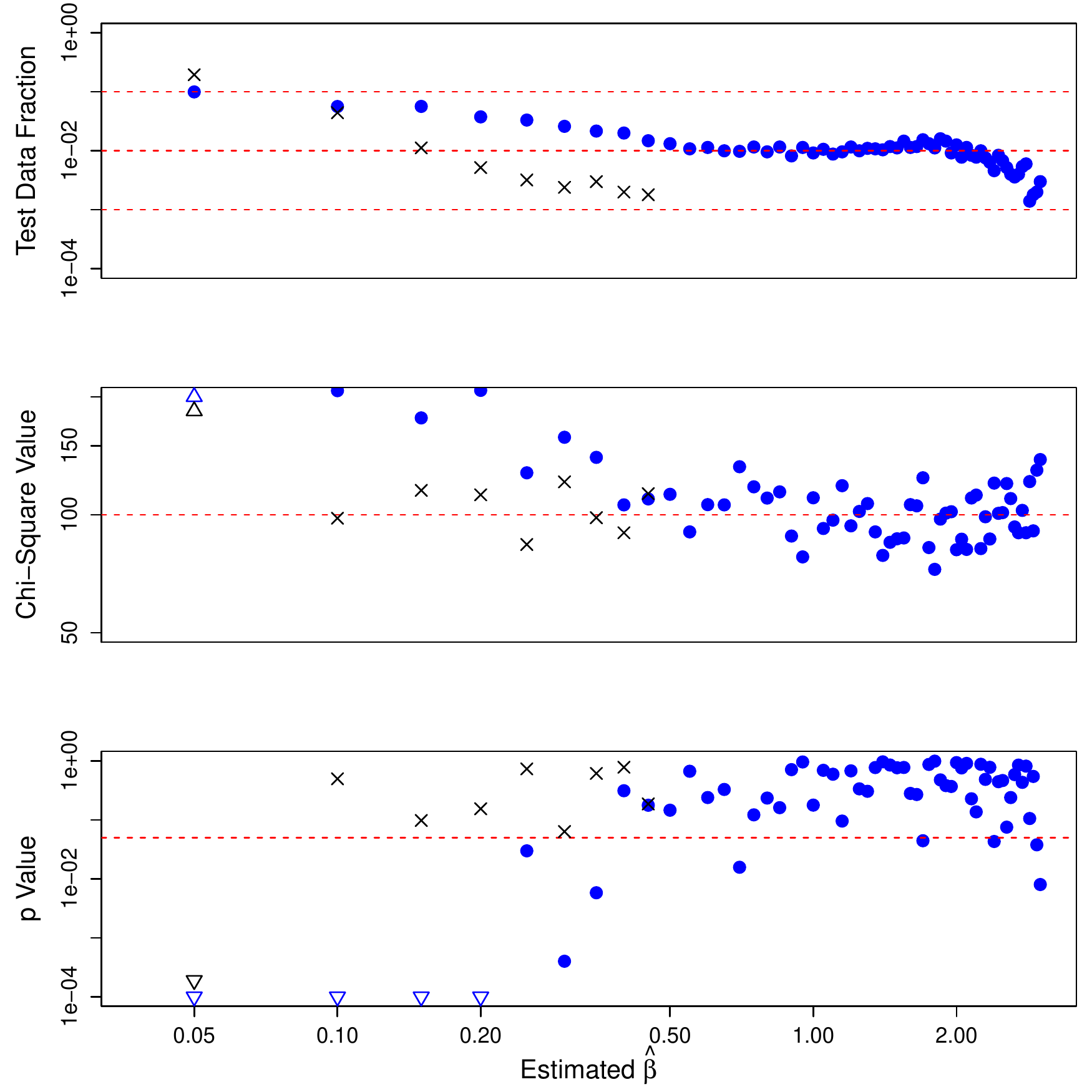}
    \caption{Results of testing whether the observed quantile value differences
    ($\widehat{c}_1-\widehat{c}_0,\ldots,\widehat{c}_{B}-\widehat{c}_{B-1}$)
    are constant across quantile bins, using the labeled test set data
    at each unique value of $\wb(\bx^L)$.
    Our estimates of $\wf(z\vert\bx)$ are generated via
    the {\tt Combined} estimator. 
    In all panels, the blue dots/black crosses represent results generated 
    with/without covariate shift mitigation (i.e.~with/without the
    incorporation of importance weights into estimation), 
    with the blue/black triangles specifying
    data points that fall outside the plot panels. 
    (There are insufficient data above
    $\wb(\bx^L) \geq 0.5$ to generate results without mitigating
    covariate shift.)
    The top panel shows the fraction of data for each unique 
    value, with the dashed red lines indicating 0.1\%, 1\%, and 10\% 
    respectively. 
    The middle panel shows the chi-square statistic 
    resulting from each test, with the red dashed
    line indicating the number of bins ($B = 100$), the value to which 
    chi square should converge if our estimator performs well
    in regions of covariate space with appreciable
    densities of unlabeled data. The bottom panel shows the $p$ value, 
    estimated via 
    simulation since the usual assumption of the chi-square goodness of
    fit test that there are at least five expected counts in each bin is
    usually violated; the red dashed line indicates $p = 0.05$, with points
    above that line indicating that the observed quantile value differences
    are consistent with constancy across quantile bins.
    The middle and bottom panels demonstrate that our method achieves desired 
    quantile-quantile behavior at $\wb(\bx^L) \gtrsim 0.3$, while also
    demonstrating the importance of assuming covariate shift with these
    data: without this assumption, we can only generate reasonable predictions 
    for unlabeled data in the regime $\wb(\bx^L) \lesssim 0.5$.
    We assume $\wb_{\rm thr} = 0.3$; analogous plots
    for $\wb_{\rm thr} = 0.1$ and $0.2$, not shown, indicate similar results.
    }
    \label{fig:qq}
\end{figure}

\begin{figure}
    \includegraphics[scale=0.475,angle=0]{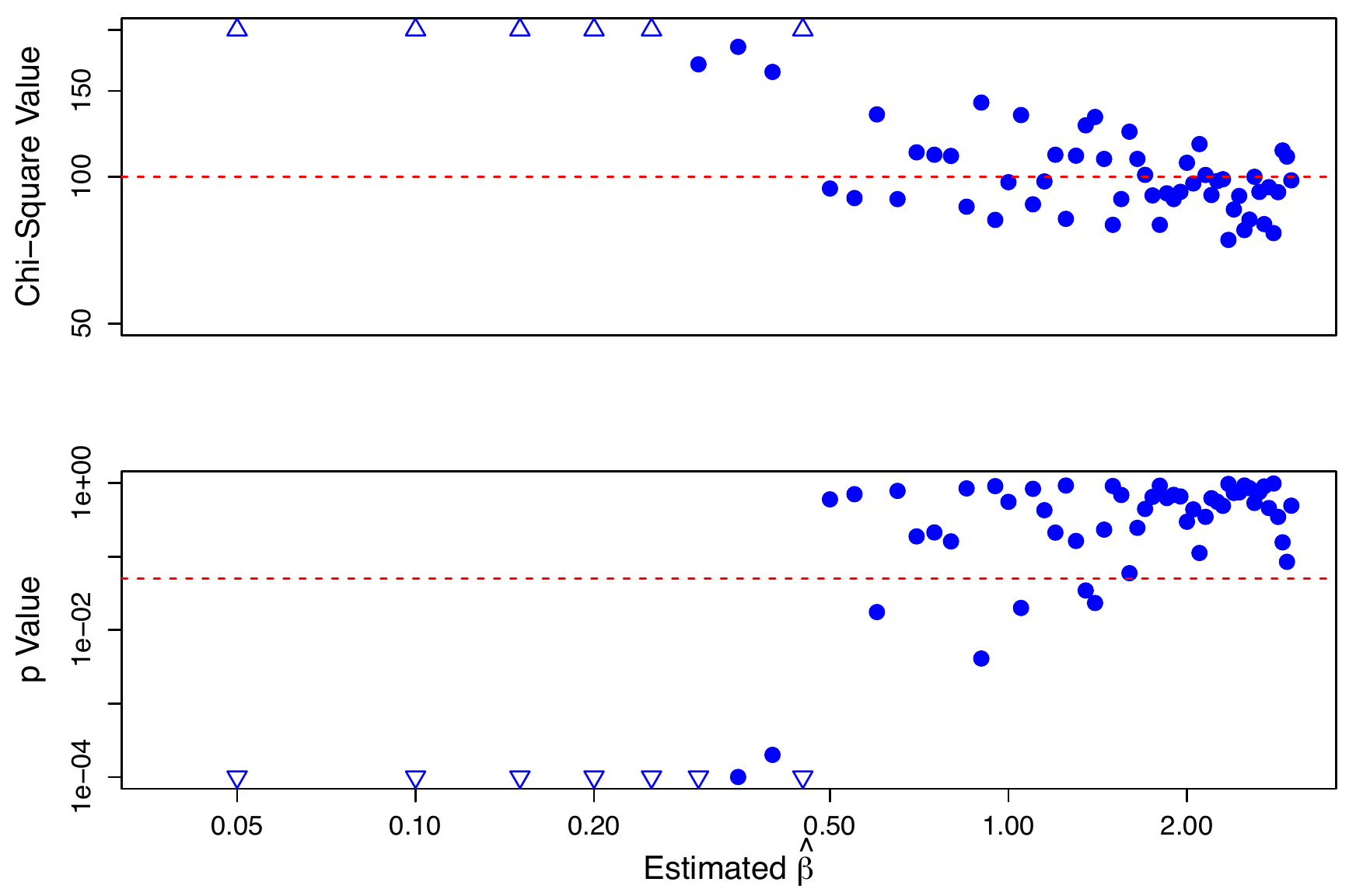}
    \caption{Same as Figure \ref{fig:qq}, except instead of quantile-quantile
    behavior we examine the differences between observed and expected
    coverage as a function of $\wb(\bx^L)$, and only for the 
    case where we mitigate covariate shift.
    The top and bottom panels demonstrate that our method 
    achieves good coverage for $\wb(\bx^L) \gtrsim 0.5$. 
    }
    \label{fig:alpha}
\end{figure}

\begin{figure}
    \includegraphics[scale=0.475,angle=0]{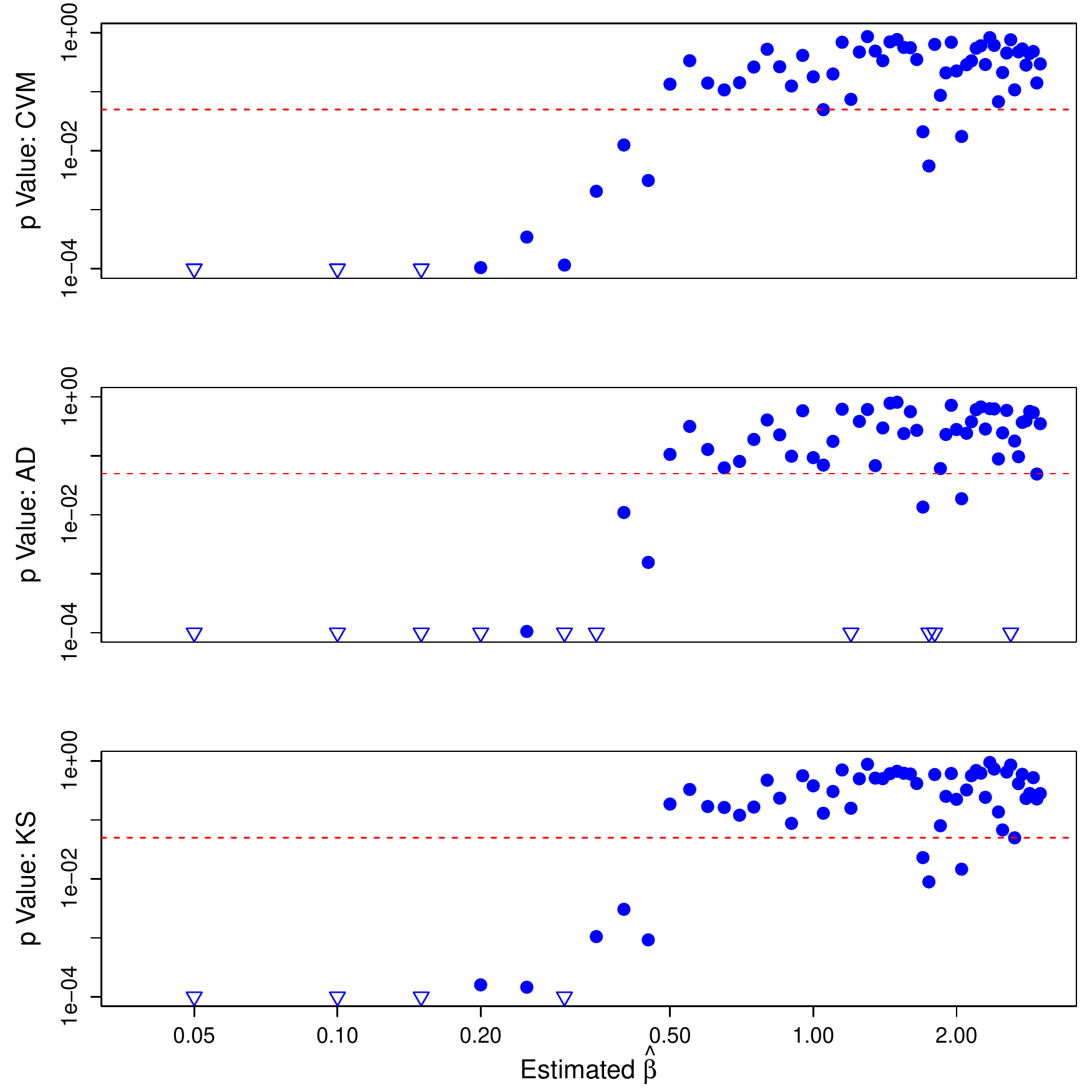}
    \caption{Results of testing whether the observed random variables 
    $U_i = \widehat{F}(z_i \vert \bx_i)$ are uniformly 
    distributed, using the labeled test set data
    at each unique value of $\wb(\bx^L)$.
    Our estimates of $\widehat{F}(z\vert\bx)$ are generated via
    the {\tt Combined} estimator. The panels show $p$ values generated via
    the Cram\'er-von Mises (CvM), Anderson-Darling (AD), and 
    Kolmogorov-Smirnov (KS) tests, from top to bottom respectively. 
    The KS test statistic is the maximum deviation of the empirical
    cumulative distribution function (CDF) for $U$ from the uniform CDF,
    while the CvM and AD statistics are based on (unweighted and weighted)
    integrated distances between the empirical and uniform CDFs.
    These panels demonstrate that our method generally 
    achieves good behavior with respect to uniformity
    at $\wb(\bx^L) \gtrsim 0.5$
    (cf.~0.3 and 0.5 for tests based on QQ and coverage plots).
    This plot shows results for the preprocessing threshold
    $\wb_{\rm thr} = 0.3$; analogous plots
    for $\wb_{\rm thr} = 0.1$ and $0.2$, not shown, indicate similar results.
    }
    \label{fig:unif}
\end{figure}

In Figure \ref{fig:qqall} we demonstrate the tradeoff that is inherent when
mitigating covariate shift, via the incorporation of importance weights
into estimation, using QQ plots. If we do not mitigate covariate
shift, we achieve good conditional density estimates within those portions 
of covariate
space in which $\wb(\bx^L) \lesssim 0.5$ (orange dashed-dotted line); however, 
$\lesssim$ 8\% of the unlabeled data reside within these portions.
Mitigation of covariate shift leads to a worsening of the CDEs within
these portions of covariate space (black dashed line), but allows one to make
good estimates throughout the remaining space (blue solid line).
In Figures \ref{fig:qq}, \ref{fig:alpha} and \ref{fig:unif} 
we show the results of 
applying hypothesis tests based on QQ plots (Section \ref{sect:qq}),
coverage plots (Section \ref{sect:cov}) and the assumption of uniformity
(Section \ref{sect:unif}).
In the first two figures we show the results of using the GoF test
to determine the consistency of expected and observed quantiles at each
unique value of $\wb(\bx^L)$, in the manner outlined
in Section \ref{sect:diag}. These results are generated using the labeled
test set data, assuming a preprocessing threshold $\wb_{\rm thr} = 0.3$.
(The optimal number of nearest neighbors given this threshold is 20, hence
the unique values of $\wb(\bx^L)$ are 0, 0.05, 0.1, etc.)
In the middle and bottom panels, we observe that for 
$\wb(\bx^L) \lesssim 0.3$, the chi-square values are much larger,
and the $p$ values much smaller, than what we would expect if $\widehat{c} = c$:
we do not achieve good behavior in this regime.
(This is consistent with the behavior of the QQ plots shown
in Figure \ref{fig:qqall}.)
For $\wb(\bx^L) \gtrsim 0.3$, on the other hand, the 
$p$ values are generally $> 0.05$.
We thus conclude that our method generates useful conditional density 
estimates in the regime $\wb(\bx^L) \gtrsim 0.3$.
(We note that we come to similar conclusions if we use the
preprocessing thresholds $\wb_{\rm thr} = 0.1$ or $0.2$ instead.)

Figure \ref{fig:unif} shows the results of applying the Cram\'er-von Mises,
Anderson-Darling, and Kolmogorov-Smirnov tests (from top to bottom,
respectively) to the data $U$ generated
from the CDFs of the conditional density estimates
$\wf(z \vert \bx)$ of the labeled test set data,
again assuming $\wb_{\rm thr} = 0.3$. We observe similar behavior for
the $p$ values here as observed in Figure \ref{fig:qq}, with two 
differences: (1) the $p$ values are generally above 0.05 for
$\wb(\bx^L) \gtrsim 0.5$ as opposed to 0.3, and
(2) there are more numerous deviations from uniformity in the regime
$\wb(\bx^L) \gtrsim 0.5$ than seen in the bottom panel
of Figure \ref{fig:qq}, particularly in the case of the AD test (middle
panel, Figure \ref{fig:unif}).

\section{Summary}

\label{sect:summary}

In this paper, we provide a principled method for generating
conditional density estimates $\wf(z\vert\bx)$
(elsewhere commonly denoted $p(z)$ and dubbed the ``photometric redshift
PDF") that takes into account selection bias and the covariate shift that
this bias induces. (See Figure \ref{fig:covshift} for an example of both: 
a bias towards brighter galaxies leads to shifted distributions of colours
between spectroscopic and photometric data samples. See also
Algorithms \ref{alg:full} and \ref{alg:preproc}.) If not mitigated,
covariate shift leads to situations where models fit to labeled 
(i.e.~spectroscopic) data will not produce scientifically useful fits 
to the far more numerous unlabeled (i.e.~photometric-only) data, lessening
the impact of photo-z estimation within the context of precision cosmology.
Here, we mitigate covariate shift by first estimating importance weights,
the ratio $\beta(\bx)$ between the densities of unlabeled and
labeled data at point $\bx$ (Section \ref{sect:iw}), 
and then applying these weights to conditional density estimates
$\wf(z\vert\bx)$ (Section \ref{sect:cde}). 

In order for our two-step procedure to succeed, ultimately, we require
good estimates of $\beta(\bx)$ at labeled data points in order
for it to achieve good estimates $\wf(z\vert\bx)$ at
unlabeled ones. We thus need both rigorously defined risk
functions that allow us to tune the free parameters of our importance weight
and conditional density estimators, and diagnostic tests that allow us
to determine the quality of the estimates 
$\wf(z\vert\bx)$.
Our method is based on the assumption that the 
probability that astronomers label
a galaxy (i.e.~determine its spectroscopic redshift) depends only on
its (photometric and perhaps other) properties $\bx$ and not 
on its true redshift, an assumption currently valid for redshifts
$\lesssim$0.5. This is equivalent to assuming that the
conditional densities for labeled and unlabeled data match
($f_L(z\vert\bx) = f_U(z\vert\bx)$), 
even if the marginal distributions differ ($f_L(\bx) \neq
f_U(\bx)$), which allows us ultimately to substitute out
the true unknown quantities $\beta(\bx)$ and
$f(z\vert\bx)$ in specifications of risk functions
(equations \ref{eqn:beta_risk} and \ref{eqn:cde_risk}).
These risk functions, and their estimates
(equations \ref{eqn:beta_risk} and \ref{eqn:cde_risk}), are the backbone
of our method: they allow us to tune parameters in a principled
manner (e.g.~what is the optimal number of nearest labeled neighbors when
estimating $\beta(\bx)$ via equation \ref{eqn:beta_nn}?),
as well as choose between competing estimators (e.g.~which is better:
the {\tt NN-CS}, {\tt kerNN-CS} or {\tt Series-CS} estimators of conditional
density?). An important question to answer in future work is whether
we can relax our central assumption and still be able to write down
estimated risks that lead to useful estimates of conditional densities in
higher redshift regimes.

In Section \ref{sect:combine}, we demonstrate that
once we generate $p$ separate conditional density estimates 
$\wf_k(z \vert \bx)$ (e.g.~via the {\tt kerNN-CS}
and {\tt Series-CS} estimators), tuned via the estimated risk given in
equation \ref{eqn:cde_risk}, we can combine them to achieve
better predictions (i.e.~smaller values of risk). The method we propose
utilizes a weighted linear combination, with the weights determinable
via quadratic programming, but this is not the only possible way to
combine estimates; see, e.g., \cite{Dahlen13}, who discuss three methods
for combining estimates, including one (Method 2) that adds estimates
together as we do, except that while we determine optimal coefficients by
minimizing estimated risk, they combine estimates so that 68.3\% of the
spectroscopic redshifts in their sample fall within their final
1$\sigma$ confidence intervals.

It is not sufficient to generate estimates $\wf(z\vert\bx)$
by minimizing risk; one also needs to demonstrate that the
estimates are scientifically useful. There is no unique way to
demonstrate the quality of conditional density estimates. In 
Section \ref{sect:diag}, we provide alternatives that test (1) whether
estimated cumulative densities, evaluated at actual redshifts, are 
distributed uniformly; (2) whether observed
quantiles match expectation via QQ plots and the chi-square GoF test; 
and (3) testing uniformity as a function of interval coverage.
The jury is still out as to which of these diagnostics
will play a central role in future photo-z analyses; for now, we consider
it sufficient to demonstrate in any analysis that these diagnostics yield 
similar qualitative results.

In Section \ref{sect:sdss}, we demonstrate our method using 
$\approx$500{,}000 galaxies with, and $\approx$500{,}000 without,
spectroscopic redshifts, mostly from the Sloan Digital Sky Survey
(see \citealt{Sheldon12} for details). For computational efficiency,
we sample 15{,}000 galaxies from both pools of data. While our initial
labeled sample is chosen randomly from the larger pool of labeled data,
we implement a preprocessing scheme (see Algorithm \ref{alg:preproc}) to 
generate a new labeled sample with a larger effective size, whose photometry
also more closely resembles that
of the unlabeled sample (see Figure \ref{fig:preproc}). The preprocessing
scheme requires the specification of an importance weight threshold
($\wb_{\rm thr}$) whose value cannot be optimized via tuning
(since different thresholds yield different labeled datasets, and thus
yield not-directly comparable estimated risks). We demonstrate that our
results are generally insensitive to the choice of threshold within the
regime $\wb_{\rm thr} \lesssim 0.3$.
Those results include
that (1) as expected, the conditional density of {\tt Combined} 
estimator, constructed from those of the {\tt kerNN-CS} and {\tt Series-CS}
estimators, provide the best estimates as quantified via the risk
estimate in equation \ref{eqn:cde_risk}, and (2) via our diagnostic tests,
we determine that our {\tt Combined} estimates exhibit good behavior in
the regimes $\wb(\bx) \gtrsim 0.3$, for QQ-based
tests, and $\gtrsim 0.5$, for tests of coverage and cumulative densities.
Our results thus demonstrate that our method achieves good,
i.e.~scientifically useful, conditional density estimates for unlabeled
galaxies.
 
\section*{Acknowledgements}

The authors would like to thank Jeff Newman (University of Pittsburgh) for
helpful discussions about photometric redshift estimation.
This work was partially supported by NSF DMS-1520786, and the National 
Institute of Mental Health grant R37MH057881.
RI further acknowledges the support of the
Funda\c{c}\~ao de Amparo \`a Pesquisa do Estado de S\~ao Paulo
(2014/25302-2).


\bsp	
\label{lastpage}
\end{document}